\documentstyle[12pt]{article}
\begin{document}
\begin{center}
\vspace*{1.0cm}
{\Large\bf A quantum model for the magnetic multi-valued recording}
\\
\vspace*{1.0cm}
  Lei Zhou, Shengyu Jin and Ning Xie\\
\vspace*{0.2cm}
{\it Department of Physics, Fudan University,
Shanghai 200433, P. R.  China}\\
\vspace*{0.6cm}
Ruibao Tao\\
\vspace*{0.2cm}
{\it Center for Theoretical Physics, Chinese Center of Advanced
Science and
Technology (World Laboratory) ,
 P. O. Box 8730,
 Beijing 100080, China }\\
{\it and Department of Physics, Fudan University,
 Shanghai 200433, P. R. China}\\
\end{center}

\vspace*{0.5cm}

\centerline{\bf\large Abstract}
\vspace*{1.0cm}

We have proposed a quantum model for the magnetic multi-valued
recording in this paper. The hysteresis loops of the two-dimensional
systems with randomly distributed magnetic atoms have been studied
by the quantum theory developed previously. The method has been proved
to be exact in this case. We find that the single-ion anisotropies and
the densities of the magnetic atoms are mainly responsible for the
hysterisis loops. Only if the magnetic atoms contained by the
systems are of different (not uniform) anistropies and their
density is low, there may be more sharp steps in the hysteresis loops.
Such materials can be used as the recording media for the so-called
magnetic multi-valued recording. Our result explained the experimental
results qualitativly.

\newpage
Magnetic thin films with perpendicular ``easy-axis" anisotropy have
attracted much attention these years both experimentally and
theoretically since such systems have good
potential to be taken as recording media \cite{mmv1}-\cite{cap}.
In practice, many efforts have been made to increase the recording
density of the devices, and several proposals have been made
to achieve this end. However, the recording density has already
come to a limit of the conventional scheme so that one must find new
approaches. Recently, experimental studies have been made on the
 magnetic multi-valued (MMV) recording which is believed to be the
next strategy of high density recording \cite{mmv1}-\cite{mmv2}.
The key point is that there should exist more metastable states
which are stable enough to record a message. Thus
the media for MMV recording must possess several sharp sub-steps
in its hysteresis loop. Experimentally, such a phenomenon had been
confirmed in some kind of magnetic layered systems \cite{mmv1}
-\cite{mmv2}. However, the theoretical origin is not yet very clear.

On theoretically side, a quantum theory for the coercive force of a
magnetic system \cite{zlprb} has been developed based on some previous
works \cite{hu1}-\cite{br}. In such a quantum approach, the concept of the
metastable state was adopted, and the magnetic excitation gap was
defined to be the order parameter to monitor the stability of such
metastable states. The coercive force can be determined by the
condition that the gap comes to zero \cite{zlprb}. The so-called
``capping effect" in a double-film structure \cite{cap} has been
explained successfully by the quantum theory \cite{zlprb}.

The present work is devoted to proposing a theorectical explanation
for the MMV recording which is confirmed by a randomly magnetic thin
film \cite{mmv2}. We first propose a model Hamiltonian for such a system
in which the magnetic atoms are distributed randomly ,
then studied its hysteresis loop by the quantum method which is proved
to be exact in such case. The results are averaged for samples
finally to overcome the fluctuations of the distribution. We show
that: 1) When the magnetic atoms contained by the system are of uniform
single-ion anisotropy, there should be only one sharp step in the
hysteresis loop.
2) When the magnetic atoms are of different anisotropies but
the density is higher than a critical value, the hysteresis
loop is highly smoothed and there are no obvious steps. 3) only when
the magnetic atoms are of different anisotropies and their density
is low, more sharp steps can be clearly observed in the hysteresis loop.

\vspace*{1.0cm}

The Hamiltonian is given as:
\begin{eqnarray}
H = \frac{1}{2}\sum_{i,j} J_{ij} {\bf S}_i \bullet {\bf S}_j
        - \sum_i D_i (S^z_i)^2  -  h\sum_i S_i^z
\end{eqnarray}
where \{ $i$ \} are the lattice sites randomly distributed in the
$x$-$y$ plane in which the magnetic atoms are occupied. $\{ D_i > 0\}$
are the single-ion anisotropies. An external field is applied along
the $z$ axis. $J_{ij} = J$ is the exchange constant and only the
nearest-neighbor interaction is considered. In real magnetic materials,
the single-ion anisotropic constant $D$ is usually much smaller than the
exchange constant $J$.

Following Refs. \cite{zlprb},\cite{br}, a local coordinates
(LC) system ($x_i,y_i,z_i$) can be introduced to the Hamiltonian.
The $x_i$ and $z_i$ axes in the LC system are rotated
by an angle $\theta_i$ which may be different from site to site, while
the $y_i$ axis is not rotated. It is helpful to apply a Bose
transformation such as Holstein-Primakoff (H-P) \cite{hp},
Dyson-Maleev (D-M) \cite{ds}-\cite{ma} and the complete
Bose transformation (CBT) \cite{cbt} to study the spin
systems. In a harmonic approximation, these transformations are the
same. However, we will use the CBT in this paper because it can present
high-order terms correctly.

After the LC transformation and the CBT, the Hamiltonian becomes:
\begin{eqnarray}
\tilde H = U_0 + H_1 + H_2 + \cdots
\end{eqnarray}
where
\begin{eqnarray}
U_0 &=& const.-\frac{1}{2} \sum_{i,j}
        J_{ij}\cos(\theta_i-\theta_j)S_iS_{j}
        -h\sum_{i}\cos\theta_iS_i\nonumber\\
    & & ~~ +\frac{1}{2}\sum_i D_{i}(2S_i-1)S_i\cos^2\theta_i\\
H_2 &=& \sum_{i,j} F_{i,j}({\theta})
         a_i^+a_j +  \sum_{i,j} G_{i,j}({\theta})(a_i^+a_j^+ + a_ia_j).
\end{eqnarray}
The coeffiecients are
\begin{eqnarray}
   F_{ii}(\theta)
&=& - D_i(S_i-\frac{1}{2})
                (\sin^2\theta_i-2\cos^2\theta_i)\nonumber\\
& &+\sum_{j}S_{j}J_{ij}\cos(\theta_i-\theta_j)
        +h\cos\theta_i,\label{eq:fm}\\
       F_{i,j}(\theta)
&=&-\frac{1}{2}J_{i,j}\sqrt{S_iS_j}
        [1+\cos(\theta_i-\theta_{j})], ~~~~~i \not= j,\\
   G_{ii}(\theta)
&=&-\frac{1}{4}\sqrt{2S_i(2S_i-1)} D_{i} \sin^2\theta_i,\\
   G_{ij}(\theta)
&=&\frac{1}{4}J_{ij}\sqrt{S_iS_{j}}
        [1-\cos(\theta_i-\theta_{j})], ~~~~i \not= j.\label{eq:gm}
\end{eqnarray}
$\{\theta_i\}$ can be obtained by minimizing the ground state energy:
$dU_0 / d\theta_i = 0$, which yield:
\begin{eqnarray}
\sum_{j}J_{ij}S_{j}
\sin(\theta_i-\theta_j) + h\sin\theta_i
+D_i(2S_i-1)\sin\theta_i\cos\theta_i = 0.\label{eq:min}
\end{eqnarray}
Equations above are the same as the condition of $H_1 = 0$. $H_2$ can be
diagonalized by a Bogolyubove transformation. So, we have:
\begin{eqnarray}
\tilde H \simeq U_0' + \sum_i \epsilon_i \alpha_i^+\alpha_i
+ \cdots \label{eq:dia}
\end{eqnarray}
In general, the nonlinear equations (\ref{eq:min}) may have many
solutions in a definate external field $h$. For each solution,
the spin state may be a {\bf metastable} one only when the excitation
energy $\{\epsilon_i\}$ based on such a solution are all positive.
In another word, the excitation should have a positive gap
$\Delta(h) > 0$, otherwise, such a spin state is not
stable. So, if every metastable states of the systems have been
investigated when one altering the external field from positive to
negative, the hysteresis loop can be obtained while the system
transits from one metastable states to another one.

Before carrying out a general investigation, we will first study
some particular systems which are illustrated in figure 1 in order
to get some informations. In figure 1 and other figures, we suppose
that the magnetic atoms represented by the circles have the anistropy
$\tilde D_1 / J = D_i(2S_i-1) / J = 0.1$, and those by the triangles
have the anistropy $\tilde D_2 / J = D_i(2S_i-1) / J = 0.02$.
It is also supposed that $S_1 = S_2$ for convenince.
It should be noted that such systems have a character that the magnetic
atoms are all coupled together since we have supposed that only the
nearest-neighour interaction is considered.

Eqs. (\ref{eq:min}) may have two kinds of solutions:
(1) The trivial solutions: \{$\theta_i=0$, or $\pi$\}.
(2) The non-trivial soltions: \{$\theta_i\not=0$, or $\pi$\}. The number
of solutions is certainly very big, and any of them may be a metastable
state if the elementary excitations have a positive gap. However, after
the numerical calculations for the systems which are illustratated in
figure 1, we find that {\bf only two} solutions among so many ones can
be metastable within a definite region of external field. They are:
(1) \{$\theta_i=0$\} (2) \{$\theta_i=\pi$\}. Other spin configurations
are all unstable. Of cause, we can not give a vigorous proof for that in a
general case, but one can understand the above numerical results as
follows. First, the ``easy-axis" of each magnetic atom is along the $z$
axis so that the spins all prefer to paralell or antiparalell the $z$
axis, and there are no such atoms which have ``in-plane easy-axis" or
``easy-plane" anisotropies, as the result, the non-trivial spin
configuration $\{\theta_i \not= 0,\pi, i=1,2,\cdots\}$ is not
likely to appear. Second, since the exchange interaction $J$ is usually
much larger than the single-ion anistropy $D$ in real material, the spins
are willing to paralell with each other. Since the spins are all coupled
together in the systems studied (figure 1), such spin configurations
that some spins are up while others are down can not be stable. This is
an explanation rather than a proof, but at least, one may understand the
numerical results that only two metastable states may exist for a
system in which the spins are coulped together.

Applying the solution \{$\theta_i=0$\} into the Hamiltonian,
we find that the excitation energy $\{\epsilon_i\}$ can be obtained by
diagonalizing the matrix $\{F_{ij}\}$ since the matrix $\{G_{ij}\}$ is
zero. In the Appendix, we prove that \{$\epsilon_l$\} are
the {\bf rigorous} excitation energies of the Hamiltonian in this
case although they seem to be obatained by a harmonic approximation.
Thus we get exactly
\begin{eqnarray}
\Delta(h) = Min[\epsilon_i],
\end{eqnarray}
and the coercive force of the model can be determined by
$\Delta(h_c)=0$.

\vspace*{1.0cm}

Following are the numerical results we obtained.

\vspace*{0.4cm}

\noindent {\bf Case 1}. Only kind of magnetic atoms are distributed in
the systems: $S_i = S$ and $(2S_i-1)D_i = \tilde D$. In this case,
the matrix $\{F_{ij}\}$ can be rewritten into the following form:
\begin{eqnarray}
\{F_{ij}\}|_{\theta_i=0} = \tilde D + h + \{ F'_{ij} \}
\end{eqnarray}
One can check at once that the matrix $\{F'_{ij}\}$ has the
lowest eigenvalue 0. Then the gap will be:
\begin{eqnarray}
\Delta(h) = \tilde D + h
\end{eqnarray}
The coercive force of the system can derived: $h_c = - \tilde D$.

Thus, no matter how the atoms are distributed in the pattern, if
there are only one kind of magnetic atoms, the coercive forces are
always the same. In practice, one makes the systems amorphous to
minimize the domain's size to improve the recording density. The above
results can warrant that every domains possess the same coercive
force so that they can be used in a same way.

\vspace*{0.4cm}

\noindent {\bf Case 2}: The system contains more than one kind of
magnetic atoms. The results for the systems shown in figure 1
are listed in Table 1,  from which we can find that the
coercive forces $h_c$ are strongly dependent on the distributions of
the magnetic atoms. Generally, the coercive force turns smaller when
the samll-anisotropy magnetic atoms are relatively increased in the system.
But the results are different for the systems in which the magnetic
atoms are coupled differently although the numbers of the two kinds
of atoms are the same.

\vspace*{1.0cm}

We will discuss the general behavious of our random model's
hysteresis loop. Throughout this paper, a $10\times 10$ lattice
is discussed just to illuminate the main physical idea, and the
samples are avaraged in the end to take account of the fluctuations.

A typical distribution is illustrated in figure 2.
One may find that there are many so-called ``Isolated Islands" (II) in
the patterns. Within each II, the magnetic atoms are coupled toghter so
that they can be treated following the method mentioned above.
Howevr, the couplying between those IIs is zero. So, such
IIs must be able to be considered independently. Actually,
the matrix ${F_{ij}}$ has the following form:
\begin{equation}
\left(
\begin{array}{ccccc}
[~~C^{(1)}~~] &               &               &           &\\
              & [~~C^{(2)}~~] &               &           &\\
              &               & [~~C^{(3)}~~] &           &\\
              &               &               & \ddots    &\\
              &               &               &           & [~~C^{(n)}~~] \\
\end{array}
\right) .
\end{equation}
where $[C^{(i)}]$ are the sub-matrixes for each IIs.
They are found not to couple with each other in the matrix.
Thus, the eigenvalues of the matrix $\{F_{ij}\}$ must be those of
the sub-matrixes $[C^{(i)}]$. Without loosing any genarality,
it can be supposed that the coercive forces $h_c$ of all the
IIs have been arranged from small to large: $|h_c^i| \leq |h_c^j|, i < j$.

The hysteresis loop of the entire structure can be obtained as follows.
As the external field $h$ reaches $h_c^1$, the minimum eigenvalue of the
sub-matrix $[C^{(1)}]$ is zero while that in others are still positive,
that means the first II is not stable while others are still stable.
It has been discussed that each II only has two metastable
configurations: the spin-up and the spin-down configuration.
So, the first II will turn to its spin-down configuration in the
vicinity of $h_c^1$ while the other IIs will remain
in there spin-up configuration. Following the same reason, the
$i$th II will turn to its spin-down configuration while other II's
configuration are not changed as the external field $h$ reaches the
$i$th coercive force $h_c^i$. If we do that step by step, we can
obtain the hysteresis loop finally. For instance, the hysteresis
loop for the system illustrated in figure 2 has been shown in
figure 3, where there are sharp decrease in the vicinites of
the coercive forces $h_c^{i}$.

However, we must do the sample avarages to overcome the fluctuations.
In our calculations, 1000 samples are averaged for a definite case.
The results are presented in figures 4-5. In the case that the density
of magnetic atoms is 40\% and the two kinds of magnetic atoms
are equally sputtered, we average two groups of 1000 smaples to
get two final results. They are compared in figure 4. However,
one may find that the difference of the two lines even can not
be detected which means the number of the samples for average
is large enough to overcome the fluctuation.
Figures 5a-5d present the hysteresis loops for the systems
in different cases. From figures 4, 5a-5c, one may find that
more sharp steps can be apperantly found in the hysteresis loop.
Futhermore, when the density is lowered, the steps are sharper and
more sub-steps in the hysteresis loops may appear (figure 5a).
This can be understood as follows. When the density is low, the
possiblities of apperaing some definite structures will be high.
As a result, there may be a distint decrease of the magnetization
in the vicinity of the coercive force for such a
structure. Actually, a limit case is that there are only two differnt
atoms in the lattice. in this case, the possibility for the distribution
that the two atoms are seperated should be greatly larger than that they
are coupled together. Thus, they may be a very clear multi-step shaped
hystersis loop.

When the density of the magnetic atoms is high, especially when the
density is larger than the percolation value, the hysteresis
loop is greatly smoothed and the step is almost undetectable (figure 5d).
Actually, near the percolation value,
the distributions of the IIs are quite complicated, and any
pattern is possible. Since the coercive force of each II is strongly
dependent on the distribution, any value of the coercive force is then
possible to appear. Thus, there should be an infinite number of
metastable states which are all different in the system.
This is very similar to that in the spin-glass system,
although such a system is quite different with that one. So, one should
not use such materials for recording.

\vspace*{0.9cm}

To summarize, in this letter, we have investigated the hysteresis loops of
the 2-d systems with randomly sputtered magnetic atoms.
The method is proved to be rigorous in this case. The results show:
a multi-step shaped hystersis loop can be achieved only if the following
two conditions can be satisfied: 1) the sputtered magnetic atoms must
have different coercive forces. 2) the density of the magnetic atoms
must be lower than the percolation value. Such materials have the
potential to be considered as the recording media for MMV recording.
If the first condition is dissatisfied,
there is only one sharp step in the hystersis loop; if the second condition,
there are no sharp step in the hystesis loop and such material can not
be used for recording.

\vspace*{1.0cm} \noindent {\bf Acknowledgments}

\vspace*{0.5cm} \noindent
This research is partly supported by National Science Foundation of China
and the National Education commission under the grant for training Ph.Ds.\\

\vspace*{2.0cm}

\noindent{\large\bf Appendix}

\vspace*{1.0cm}

\noindent In this appendix, we will prove that $\{\epsilon_i\}$
are the exact excitation energies of the Hamiltonian (1) in the
case of $\theta_i=0$.

Since $\theta_i=0$, the Bose transformation can be applied naively to
Hamiltonian (1). Following Ref. \cite{cbt}, the CBT is given as:
\begin{equation}
\left\{
\begin{array}{ll}
\displaystyle S_i^z=\sum_{l=0}^{\infty} A_l^i a_i^{+ l} a_i^l\\
\displaystyle S^+_i=\sum_{l=0}^{\infty} B_l^i a_i^{+ l} a_i^{l+1}\\
\displaystyle S^-_i=\sum_{l=0}^{\infty} B_l^i a_i^{+ l+1} a_i^{l}\\
\end{array}
\right.
\end{equation}
$\{ A_l^{i}\}$,$\{ B_l^{i}\}$ are the coefficients of Bose expansion
which are dependent on $S$. We can also derive the
expansions for the single-ion anisotropy terms:
\begin{eqnarray}
(S_i^z)^2 = \sum_{l=0}^{\infty} G_l^i a_i^{+l}a_i^l
\end{eqnarray}

Applying the CBT to the Hamiltonian, we have a transformed Hamiltonian
$\tilde H$ which has $\bf exactly$ the same eigenvalues as the
Hamiltonian (1):
\begin{eqnarray}
\tilde H = U_0 + H_2 + H_4 + \cdots.
\end{eqnarray}
where
\begin{eqnarray}
H_2 &=& \sum_{i,j} F_{ij} a_i^+ a_j\\
\cdots
\end{eqnarray}

Of course, Hamiltonian $\tilde H$ is still impossible to solve exactly.
However, some eigemstates can be obtained exactly.
$H_2$ can be diagonalized by a orthogonal
transformation:
\begin{eqnarray}
\tilde a^+_m = \sum_n P_{mn} a^+_n
\end{eqnarray}
After the transformation, we find
\begin{eqnarray}
H_2 = \sum_n \epsilon_n \tilde a_n^+ \tilde a_n
\end{eqnarray}
where $\{ \epsilon_n \}$ are the eigenvalues of the matrix $\{F_{ij}\}$
and $P_{mn}$ can be found from calculating the eigenvectors of the
matrix $\{F_{ij}\}$.

From the CBT, we find that every terms in the remainder interaction
$H_I=H_4 + H_6 + \cdots$ contains equivalent numbers of the creation
operators $a^+$ and annihilation operators $a$ and contains at least
two annihilation operators. For example, $H_4$ is found to be
\begin{eqnarray}
H_4 &=&\frac{1}{2}\sum_{ij}J_{ij}[A_0^iA_2^ia_i^{+2}a_i^2
+A_0^jA_2^ja_j^{+2}a_j^2 + A_1^jA_1^ja^+_ia^+_ja_ia_j\nonumber\\
& &~~~~+ C_1^iC_0^ja^+_ia^+_j(a^2_i + a_j^2)]
-\sum_i D_i G_2^i a^{+2}_ia^2_i - h\sum_i A_2^ia^{+2}_ia_i^2.
\end{eqnarray}
Then, it is easy to prove that
\begin{eqnarray}
H_I \tilde a_n^+|0\rangle = \sum_m P_{n,m} H_I a_m^+|0\rangle \equiv 0
\end{eqnarray}
As the result,
\begin{eqnarray}
\tilde H (\tilde a_n^+|0\rangle) = \epsilon_n (\tilde a_n^+|0\rangle)
\end{eqnarray}
So, such one-magnon eigenstates $\tilde a_n|0\rangle$ are the
{\bf exact} eigenstates of the system,and $\{\epsilon_i\}$
must be the {\bf exact} excitation energies of the system.

\vspace*{2.0cm}

\noindent {\large\bf Captions:}

\vspace*{0.7cm}

\noindent Figure 1: 
\parbox[t]{12cm} {Some particulr pattterns studied in this paper} \\

\noindent Figure 2: 
\parbox[t]{12cm}{A typical distribution of the magnetic ions in the
system we sudied} \\

\noindent Figure 3: 
\parbox[t]{12cm}{The hystersis loop of the system illustrated in
figure 2. } \\

\noindent Figure 4: 
\parbox[t]{12cm}{Comparison of the hysteresis loops obtained by
averaging two groups of 1000 samples seperately}\\

\noindent Figure 5:
\parbox[t]{12cm}{Hysteresis loops for the systems in different cases}\\

\begin{table}
\caption{Coercive forces $h_c$ for some particular patterns illustarted
in figure 1}
\begin{center}
\begin{tabular}{|c|c|}
\hline\hline
{\sl Pattern}& {\sl Coercive forces $h_c$}\\
 \hline
A    &       0.059200319744255659  \\
B    &       0.046196842386699633  \\
C    &       0.045505418405692213  \\
D    &       0.039408095183044223  \\
E    &       0.038938721535082826  \\
F    &       0.079256330941564368  \\
\hline\hline
     \end{tabular}
  \end{center}
\end{table}


\vspace*{2.0cm}

\begin{thebibliography}{99}
\vspace*{1.0cm}
\bibitem{mmv1}  K. Shimazaki, M. Yoshihiro, O. Ishizaki, S. Ohnuki
                and N. Ohta, J. Magn. Soc. No. S1, 429 (1995).
\bibitem{mmv2}  S. Gadetsky, T. Suzuki, J. K. Erwin and M. Mansuripur,
                J. Magn. Soc. No. S1, 91 (1995).
\bibitem{jpc}   R E Camley and R L Stamps, J. Phys.: Condens. Matter
                {\bf 5} 3727 (1993).
\bibitem{cap}   S. Ohnuki, K. Shimazaki, N. Ohta and H. Fujiwara,
                J. Magn. Soc. Jpn, {\bf 15} Supplement No. S1,
                399 (1991).
\bibitem{zlprb}    Lei Zhou and Ruibao Tao, ``Quantum theory of the
                coercive  force and the capping effect for magnetic
                multilayer" Phys. Rev. {\bf B}. (1996) to be
                published in the 01Oct issue.
\bibitem{hu1}    X. Hu and Kawazoe, J. Appl. Phys., {\bf 75} 6486 (1994).
\bibitem{hu2}    X. Hu and Y. Kawazoe,
                Phys. Rev. {\bf B 49} 3294 (1994).
\bibitem{br}    Ruibao. Tao, Xiao Hu and Yoshiyuki Kawazoe,
              Phys. Rev. {\bf B 52} 6178 (1995).
\bibitem{hp}    T. Holstein and H. Primakoff, Phys. Rev {\bf 59} 1098
              (1940).
\bibitem{ds}    F. J. Dyson, Phys. Rev. {\bf 102}, 1217 (1956).
\bibitem{ma}    S. V. Maleev, Zh. Eksp. Teor. Fiz. {\bf 30}, 1010
                (1957).
\bibitem{cbt}    Ruibao Tao, J. Phys. A: Math. and  Gen. {\bf 27} 3621
                (1994).
\end{thebibliography}
\end{document}